\documentclass[11pt]{article}

\usepackage{fullpage}

\usepackage{amsmath,amsfonts,amsthm,amssymb,graphicx,epic,eepic,multicol}

\theoremstyle{definition}
\newtheorem{definition}{Definition}
\theoremstyle{remark}
\newtheorem{remark}[definition]{Remark}

\newtheoremstyle{mytheorem}{0.5cm}{0.2cm}{\slshape}{ }{\bfseries}{.}{ }{}
\theoremstyle{mytheorem}

\newtheorem{prop}[definition]{Proposition}

\newcommand{\E}{\mathbf{E}}
\newcommand{\R}{\mathbb{R}}
\DeclareMathOperator{\one}{{1\hspace*{-0.55ex}I}}

\newcommand{\Q}{\mathbf{Q}}

\renewcommand{\phi}{\varphi}

\newcommand{\thf}{\frac{1}{2}\,}

\newlength{\querylen}
\usepackage{fancybox}

\numberwithin{equation}{section}
\numberwithin{definition}{section}

\begin{document}

\title{Static replications with traffic light options}

\author{Michael Schmutz\thanks{Department of
    Mathematical Statistics and Actuarial Science, University of Bern,
    Sidlerstrasse 5, 3012 Bern, Switzerland ({\tt
    michael.schmutz@stat.unibe.ch})}\ \,
    and Thomas Z\"urcher\thanks{Department of Mathematics and
    Statistics, P.O.~Box 35 (MaD), 40014 University of Jyv\"askyl\"a,
    Finland and Mathematical Institute, University of Bern,
    Sidlerstrasse 5, 3012 Bern, Switzerland
    ({\tt thomas.zuercher@gmail.com})}}

\maketitle

\begin{abstract}
  It is well known that any sufficiently regular one-dimensional
  payoff function has an explicit static hedge by bonds, forward
  contracts and lots of vanilla options. We show that the natural
  extension of the corresponding representation leads to a static
  hedge based on the same instruments along with traffic light
  options, which have recently been introduced in the market.
  One big advantage of these replication strategies is
  the easy structure of the hedge. Hence,
  traffic light options are particularly powerful building
  blocks for more complicated bivariate options. While
  it is well known that the second strike derivative of
  non-discounted prices of vanilla options are related to the
  risk-neutral density of the underlying asset price in the corresponding
  absolutely continuous settings, similar statements hold for traffic
  light options in sufficiently regular bivariate settings.

  \medskip


\end{abstract}

\section{Introduction}
\label{sec:introduction}

Particularly in Europe, there has been a liquid market in
structured products during recent years. At present, the majority
of the trades still occur over the counter, but more and more
trades are also organised at exchanges, especially at the quite
new European exchange for structured products, Scoach. Quite
often, structured products depend on more than one underlying
asset. The increasing importance of quite complicated financial
derivatives written on two or more underlying asset prices is also
reflected in the fast growing literature about multivariate
valuation technics, see e.g.~the very recent article by Eberlein
et al.~\cite{ebe:gla:pap10} and the literature cited therein. Also
quite recently, several London-based investment banks have
independently developed so-called traffic light options. In their
purest form, traffic light options are the product of a standard
equity put option and an interest rate floorlet. These products
have been developed to suit the needs of Danish life and pension
companies, see e.g.\ the paper by J{\o}rgensen~\cite{jor07}. More
generally, traffic light options are European options with payoffs
of the form
\begin{equation}
  \label{eq:tlo}
  g_{k_1,k_2}(S_{T1},S_{T2})=
  \begin{cases}
  (k_1-S_{T1})_+(k_2-S_{T2})_+\,,\\
  (k_1-S_{T1})_+(S_{T2}-k_2)_+\,,
  \\ (S_{T1}-k_1)_+(k_2-S_{T2})_+\,,\\
  (S_{T1}-k_1)_+(S_{T2}-k_2)_+\,,
  \end{cases}
\end{equation}
where $S_{Ti}$, $i=1,2$, stand for two asset prices at maturity
$T>0$, $k_1$, $k_2\geq 0$ are fixed strike prices, and
$(a)_+=\max(a,0)$ for $a\in\R$. Later, we will also use the
notation $\R_+=[0,\infty)$. In view of their increasing popularity
in practice, also the theoretical interest in these products
increased recently, see e.g.~\cite{jor07,kok09,pou09}. Depending
on the author, the same, or closely related options, are also
called \emph{correlation options} and have been analysed in
relation to multivariate valuation techniques, see
e.g.~\cite{bas:mad00,dem:hon02,ebe:gla:pap10,kwo:leu:won10,lee:04,zan}.

The theoretical popularity of these options is mainly based on the
fact that they are relatively easy to price, while Bakshi and
Madan~\cite{bas:mad00} also remarked that the product of two calls
on the price change factors is market completing in the sense
of~\cite{nac88,ros76}. In this note, we emphasize two further,
quite general but still explicit properties, being related to the
latter one, which are analogues of very important and frequently
used theoretical properties of vanilla options in the
one-dimensional case. First, we show that in sufficiently regular
absolutely continuous settings, the (implicit) joint risk-neutral
density is directly related to the (non-discounted) prices of
traffic light options, yielding a natural extension of a famous
result due to Breeden and Litzenberger~\cite{bre:lit78}, namely
that in sufficiently regular one-dimensional cases, the second
strike derivatives of (non-discounted) prices of European vanilla
options correspond to the risk-neutral probability density
function. In view of the results
in~\cite{car:laur09,hen:sha90,lip01}, it turns out that traffic
light options are more directly related to the (implicit) joint
risk-neutral density than basket options. Furthermore, we show (in
a purely analytical way, i.e.\ without having a probability space
in the background) that traffic light options also have a great
potential as building blocks for explicit static hedges.
Especially in view of the observation that static hedges based on
basket options seem to be harder to derive explicitly, c.f.\
e.g.~\cite{bax98,hen:sha90,lip01} for certain families of
functions, the easy structure of the resulting hedges seems to be
a substantial advantage of traffic light options.

To sum up, traffic light options are quite easy to price, they
directly reflect the joint risk-neutral distribution, and they are
also efficient building blocks for static hedges, i.e.\ there are
in fact important reasons to appreciate the market introduction of
these products.

\section{Traffic light options and the risk-neutral density}

The observations concerning the probability density are quite
obvious, but to our best knowledge, have not yet been reported in
the existing literature.

\begin{prop}
  \label{prop:density}
  Assume a risk-neutral absolutely continuous setting with continuous Lebesgue
  density denoted by \mbox{$q\colon\R_+^2\to\R_+$}. Then for $k_1,k_2>0$
  \begin{displaymath}
    q(k_1,k_2)=\frac{\partial^4}{\partial k_1\partial k_2\partial k_1\partial k_2}
    (\E[(k_1-S_{T1})_+(k_2-S_{T2})_+])\,.
  \end{displaymath}
\end{prop}

\begin{proof}
  Note that the expected payoffs
  in Proposition~\ref{prop:density} are finite. Furthermore,
  \begin{align*}
    \E[(k_1-S_{T1})_+(k_2-S_{T2})_+]&=\int_{\R_+^2}(k_1-x)_+(k_2-y)_+q(x,y)dxdy\\
    &=\int_0^{k_2}\int_0^{k_1}(k_1-x)(k_2-y)q(x,y)dxdy\,.
  \end{align*}
Differentiating yields the result.
\end{proof}

Note that in the above setting the (implicit risk-neutral)
cumulative distribution function is obtained by
\begin{displaymath}
    \Q(S_{T1}\leq k_1,S_{T_2}\leq k_2)=\frac{\partial^2}{\partial k_1\partial
    k_2}(\E[(k_1-S_{T1})_+(k_2-S_{T2})_+])\,.
\end{displaymath}

\begin{remark}[Other traffic light and correlation options]
  Note that the integrals in the proof of Proposition~\ref{prop:density}
  are only non-vanishing on bounded sets. In order to avoid problems
  being caused by integrating over non-bounded intervals in other
  cases, we impose some more regularity. More concretely, we assume
  that besides of the existence of a continuous joint density that the
  marginal distributions also exhibit continuous densities and that
  $\int_{\R_+}xq(x,y)\,dx$ (respectively $\int_{\R_+}yq(x,y)\,dy$)
  is continuous in $y$ (respectively $x$).
  Furthermore, we need $\E[S_{T1}S_{T2}]<\infty$ (along with $\E [S_{T1}]$,
  $\E [S_{T2}]<\infty$, imposing no extra restriction in our
  risk-neutral setting). Then, again by suitably writing down the expectations
  similarly as above and by differentiating, we obtain for
  $k_1,k_2>0$
  \begin{displaymath}
    q(k_1,k_2)=\frac{\partial^4}{\partial k_1\partial k_2\partial k_1\partial k_2}
    (\E[g_{k_1,k_2}(S_{T1},S_{T2})])\,,
  \end{displaymath}
  with $g$ being either one of the functions given in~(\ref{eq:tlo}).

  As already mentioned, traffic light options are sometimes also
  called correlation options. However, more usually, correlation
  options represent options of the type
  \begin{displaymath}
   (S_{T1}-k_1)_+\one_{S_{T2}>k_2}\,,\quad
   (k_1-S_{T1})_+\one_{S_{T2}<k_2}\,.
  \end{displaymath}
  It almost goes without saying that the densities are also
  implicit in the prices of these options under the imposed assumptions.
  More concretely, differentiating the (non-discounted) prices (for arbitrary
  combinations of strikes) of
  these products two times with respect to $k_1$ and once with
  respect to $k_2$ yields (for the first product up to a minus sign)
  the joint risk-neutral probability density (in a corresponding absolutely
  continuous setting). Quite closely related to these observations is the well-known fact,
  see e.g.~\cite{bru,cher:luc:vec04}, that the prices of certain
  bivariate digital options also directly reflect the risk-neutral
  distribution. For example the (implicit) joint risk-neutral bivariate cumulative
  distribution function is given by $\Q(S_{T1}\leq k_1,S_{T2}\leq k_2)
  =\E_\Q[\one_{S_{T1}\leq k_1}\one_{S_{T2}\leq k_2}]$, i.e.\ by the
  non-discounted prices of a certain bivariate binary put (with arbitrary
  strike combinations).
\end{remark}

Hence, theoretically, the non-discounted prices of a wide variety
of traffic light options (or correlation options) easily yield the
joint risk-neutral density. However, as in the one-dimensional
setting, one has to keep in mind that since these expressions
involve derivatives of (incomplete) market data, the canonical
strategy leads to ill-posed problems and regularization will be
needed.

\section{Traffic light options in static hedges}
  \label{sec:bivar-payoff}

In what follows, we use the convention
$\int_a^bf(x)dx=-\int_b^af(x)dx$. Furthermore, for $A\subset\R^n$,
we say that $f\colon A\to\R$ is differentiable of a certain order
on $A$ if $f$ can be extended to a \emph{differentiable function
of the same order} on an open set $U\supset A$. We start with a
short discussion of a well-known univariate result, due to Carr
and Madan~\cite{car:mad94}. Assume, as in Bakshi and
Madan~\cite{bas:mad00}, that a payoff function $f\colon\R_+\to \R$
is two times continuously differentiable (not necessarily
integrable). As e.g.\ in~\cite{lip01}, by the fundamental theorem
of calculus, by integration by parts, and by the formula
$xf'(x)=\int_a^xxf''(t)dt+xf'(a)$, we have, for any $a\in\R_+$
\begin{align}
  f(x)&=f(a)+\int_a^x f'(k)dk=f(a)+xf'(x)-af'(a)-\int_a^x kf''(k)dk\nonumber\\
  &=f(a)+\int_a^xxf''(k)dk+xf'(a)-af'(a)-\int_a^x kf''(k)dk\nonumber\\
  &=f(a)+f'(a)(x-a)+\int_a^xf''(k)(x-k)dk\nonumber\\
  &=f(a)+f'(a)(x-a)+\int_a^xf''(k)(x-k)_+dk-\int_a^xf''(k)(k-x)_+dk\nonumber\,.
\end{align}
We arrive at
\begin{equation}
  \label{eq:univar-rep}
  f(x)=f(a)+f'(a)(x-a)+\int_a^\infty f''(k)(x-k)_+dk+\int_0^a f''(k)(k-x)_+dk\,,
\end{equation}
$x\in\R_+$, a well-known representation, see
e.g.~\cite{bas:mad00,car:cho97,car:cho02,car:mad94,hen,lip01}. The
different original proof is presented in~\cite{car:mad94}. The
economical interpretation of this representation is that if we let
$a$ be the current forward price, we have a static hedge with
bonds, forwards, and lots of options (where the options are out of
and at the money in a certain sense).

In the bivariate setting, we start by considering four times
continuously differentiable (not necessarily integrable) payoff
functions $f\colon\R_+^2\to\R$. We introduce some efficient
notation. By $f_{i}$ we denote the partial derivative with respect
to the $i$th component, by $f_{ij}$ the second partial derivative
calculated first with respect to the $i$th then with respect to
the $j$th component, etc.

Similarly, as in the derivation of~(\ref{eq:univar-rep}), we have
for any fixed $(a,b)\in\R_+^2$, $(x,y)\in\R_+^2$
\begin{displaymath}
  f(x,y)=-f(a,b)+f(x,b)+f(a,y)+\int_a^x\int_b^y
  f_{12}(k_1,k_2)dk_2dk_1\,.
\end{displaymath}
By applying~(\ref{eq:univar-rep}) for $f(x,b)$ and $f(a,y)$, we
arrive at
\begin{displaymath}
  f(x,y)=I_1+I_2+I_3\,,
\end{displaymath}
where
\begin{align*}
  I_1&=f(a,b)+f_1(a,b)(x-a)+f_2(a,b)(y-b)+\int_0^a f_{11}(k_1,b)(k_1-x)_+dk_1\\
  I_2&=\int_a^\infty f_{11}(k_1,b)(x-k_1)_+dk_1+\int_0^b f_{22}(a,k_2)(k_2-y)_+dk_2+\int_b^\infty f_{22}(a,k_2)(y-k_2)_+dk_2
\end{align*}
and
\begin{align*}
  I_3&=\int_a^x\int_b^y f_{12}(k_1,k_2)dk_2dk_1\nonumber\\
  &=\int_a^x\left(yf_{12}(k_1,y)-bf_{12}(k_1,b)-\int_b^y k_2f_{122}(k_1,k_2)dk_2\right)dk_1\nonumber\\
  &=\int_a^x\left(\int_b^y yf_{122}(k_1,k_2)dk_2+yf_{12}(k_1,b)-bf_{12}(k_1,b)-\int_b^yk_2f_{122}(k_1,k_2)dk_2\right)dk_1\nonumber\\
  &=\int_a^x \left(f_{12}(k_1,b)(y-b)+\int_b^y(y-k_2)f_{122}(k_1,k_2)dk_2\right)dk_1\nonumber\\
  &=\int_a^xf_{12}(k_1,b)(y-b)dk_1
    +\int_a^x\int_b^yf_{122}(k_1,k_2)(y-k_2)dk_2dk_1=I_{31}+I_{32}\,.
\end{align*}
The last two summands can be written as
\begin{align*}
  &I_{31}=(y-b)\int_a^xf_{12}(k_1,b)dk_1\\
  &=(y-b)\big(xf_{12}(x,b)-af_{12}(a,b)-\int_a^xk_1f_{121}(k_1,b)dk_1\big)\\
  &=(y-b)\big(\int_a^x xf_{121}(k_1,b)dk_1+xf_{12}(a,b)-af_{12}(a,b)-\int_a^xk_1f_{121}(k_1,b)dk_1\big)\\
  &=(y-b)\big(f_{12}(a,b)(x-a)+\int_a^xf_{121}(k_1,b)(x-k_1)dk_1\big)\\
  &=(y-b)\big(f_{12}(a,b)(x-a)+\int\limits_a^xf_{121}(k_1,b)(x-k_1)_+dk_1
          -\int\limits_a^x f_{121}(k_1,b)(k_1-x)_+dk_1\big)\\
  &=f_{12}(a,b)\big((x-a)_+(y-b)_+-(x-a)_+(b-y)_+-(a-x)_+(y-b)_++(a-x)_+(b-y)_+\big)\\
  &\qquad\qquad\qquad+\int\limits_a^\infty f_{121}(k_1,b)\big((y-b)_+-(b-y)_+\big)(x-k_1)_+dk_1\\
  &\qquad\qquad\qquad\qquad\qquad\qquad\qquad\qquad\qquad
   +\int\limits_0^a
   f_{121}(k_1,b)\big((y-b)_+-(b-y)_+\big)(k_1-x)_+dk_1\,,
\end{align*}
and, by applying Fubini's Theorem, along with the same arguments
\begin{align*}
  I_{32}&=\int_a^x\int_b^y f_{122}(k_1,k_2)(y-k_2)dk_2dk_1=\int_b^y(y-k_2)\int_a^xf_{122}(k_1,k_2)dk_1dk_2\\
  &=\int_b^y(y-k_2)\left(f_{122}(a,k_2)(x-a)+\int_a^x f_{1221}(k_1,k_2)(x-k_1)dk_1\right)dk_2\\
  &=(x-a)\int_b^yf_{122}(a,k_2)(y-k_2)dk_2
  +\int_b^y\int_a^x f_{1221}(k_1,k_2)(y-k_2)(x-k_1)dk_1dk_2\\
  &=\int_b^\infty
  f_{122}(a,k_2)\big((x-a)_+-(a-x)_+\big)(y-k_2)_+dk_2\\
  &\qquad+\int_0^bf_{122}(a,k_2)\big((x-a)_+-(a-x)_+\big)(k_2-y)_+dk_2\\
  &\qquad\qquad\qquad\qquad\qquad\qquad\qquad\qquad
    +\int_b^y\int_a^x f_{1221}(k_1,k_2)(y-k_2)(x-k_1)dk_1dk_2\,.
\end{align*}
Finally, we rewrite the last summand as
\begin{align*}
  \int_b^y\int_a^x f_{1221}&(k_1,k_2)\big((y-k_2)_+-(k_2-y)_+\big)
     \big((x-k_1)_+-(k_1-x)_+\big)dk_1dk_2\\
  &=\int_b^\infty\int_a^\infty f_{1221}(k_1,k_2)(x-k_1)_+(y-k_2)_+dk_1dk_2\\
  &\  +\int_b^\infty\int_0^a f_{1221}(k_1,k_2)(k_1-x)_+(y-k_2)_+dk_1dk_2\\
  &\  +\int_0^b\int_a^\infty f_{1221}(k_1,k_2)(x-k_1)_+(k_2-y)_+dk_1dk_2\\
  &\  +\int_0^b\int_0^a f_{1221}(k_1,k_2)(k_1-x)_+(k_2-y)_+dk_1dk_2\,.
\end{align*}

To sum up, we get for $(x,y)\in\R_+^2$
\begin{eqnarray}
  \label{eq:long-version}
  f(x,y) &= &f(a,b)+f_1(a,b)(x-a)+f_2(a,b)(y-b)\\
 & &+f_{12}(a,b)((x-a)_+(y-b)_++(a-x)_+(b-y)_+) \nonumber \\
 & &-f_{12}(a,b)((x-a)_+(b-y)_++(a-x)_+(y-b)_+)\nonumber \\
 & &+\int_0^a\left( f_{11}(k_1,b)(k_1-x)_++f_{121}(k_1,b)\big((y-b)_+-(b-y)_+\big)(k_1-x)_+\right)dk_1\nonumber \\
 & &+\int_a^\infty \left(f_{11}(k_1,b)(x-k_1)_++f_{121}(k_1,b)\big((y-b)_+-(b-y)_+\big)(x-k_1)_+\right)dk_1\nonumber\\
 & &+\int_0^b \left(f_{22}(a,k_2)(k_2-y)_++f_{122}(a,k_2)\big((x-a)_+-(a-x)_+\big)(k_2-y)_+\right)dk_2\nonumber \\
 &&+\int_b^\infty \left(f_{22}(a,k_2)(y-k_2)_++f_{122}(a,k_2)\big((x-a)_+-(a-x)_+\big)(y-k_2)_+\right)dk_2\nonumber
 \end{eqnarray}\vspace{-10mm}
 \begin{align*}
 &  +\int_0^b \left(\int_0^a f_{1221}(k_1,k_2)(k_1-x)_+(k_2-y)_+dk_1+\int_a^\infty f_{1221}(k_1,k_2)(x-k_1)_+(k_2-y)_+dk_1\right)dk_2\\
 &  +\int_b^\infty \left(\int_0^a f_{1221}(k_1,k_2)(k_1-x)_+(y-k_2)_+dk_1+\int_a^\infty f_{1221}(k_1,k_2)(x-k_1)_+(y-k_2)_+dk_1\right)dk_2\,.
\end{align*}

The economical interpretation of this representation is that if we
let $a$ be the current forward price of the first and $b$ the one
of the second asset, we have a static hedge with bonds, forwards,
and again lots of options, namely vanilla and traffic light
options (being out of and at the money in a certain sense).
However, the hedging formula simplifies considerably for $a=b=0$
\begin{align}
  f(x,y)&=f(0,0)+f_1(0,0)x+f_2(0,0)y+f_{12}(0,0)xy\nonumber\\
        &+\int_0^\infty
        f_{11}(k_1,0)(x-k_1)_++f_{121}(k_1,0)y(x-k_1)_+dk_1\nonumber\\
        &+\int_0^\infty
        f_{22}(0,k_2)(y-k_2)_++f_{122}(0,k_2)x(y-k_2)_+dk_2\nonumber\\
        \label{eq:00}
        &+\int_0^\infty\int_0^\infty
        f_{1221}(k_1,k_2)(x-k_1)_+(y-k_2)_+dk_1dk_2\,,
\end{align}
where the underlying possible asset prices $x$ and $y$ can also be
interpreted as payoff of zero-strike options. This representation
simplifies considerably for several concrete examples. E.g.\ if we
would like to replicate some other well-known building blocks for
certain families of functions or for approximations, we e.g.\
obtain for $(x,y)\in\R_+^2$
\begin{align*}
  xy^2&=\int_0^\infty 2x(y-k_2)_+dk_2\,,\\
  x^iy^j&=\int_0^\infty\int_0^\infty
  i(i-1)j(j-1)k_1^{i-2}k_2^{j-2}(x-k_1)_+(y-k_2)_+dk_1dk_2\,,\quad\text{where } i,j\geq
  2,
\end{align*}
or for $h(x,y)=\frac{1}{2\pi}\exp(-\thf(x^2+y^2))$,
\begin{multline*}
  h(x,y)=\frac{1}{2\pi}\Big(1+\int_0^\infty(k_1^2-1)e^{-\thf
  k_1^2}(x-k_1)_+dk_1+\int_0^\infty(k_2^2-1)e^{-\thf
  k_2^2}(y-k_2)_+dk_2\\
  +\int_0^\infty\int_0^\infty(1+k_1^2k_2^2-k_1^2-k_2^2)
  e^{-\thf(k_1^2+k_2^2)}(x-k_1)_+(y-k_2)_+dk_1dk_2\Big)\,.
\end{multline*}
It is stressed e.g.\ by Delbaen and Schachermayer~\cite{del:sch05}
that a general analysis of financial markets should also consider
situations where prices, at least for some instruments, can be
negative. In view of that, it could be worth noticing that an only
minimally modified version of the
representation~(\ref{eq:long-version}) also holds for four times
continuously differentiable (not necessarily integrable) functions
$f\colon\R^2\to\R$. The modification is obtained by allowing
$a,b\in\R$ and by changing the integral limits from $0$ to
$-\infty$ everywhere, where the limits are $0$
in~(\ref{eq:long-version}). The steps in the proof remain
unchanged. However, note that the corresponding version
of~(\ref{eq:00}) and of the subsequent examples for general
$(x,y)\in\R^2$ remain slightly more complicated, since it is a
priori not clear which terms in~(\ref{eq:long-version}) vanish
(this depends on the sign of $x$ and $y$). Furthermore, observe
that the translates of $h$ as function on $\R^2$ are four times
continuously differentiable and thus, can theoretically (i.e.\ the
availability of the hedging instruments is assumed) be represented
with the modified representation, while at the same time it is
well known that the Fourier transform of $h$ never vanishes.
Hence, very similarly as in Bakshi and Madan~\cite{bas:mad00}, it
follows from some extended (since here we are in the bivariate
case) versions of Wiener's Approximation Theorem that based
on~\cite[Cor.~7.2.5d]{rud} (or even more directly based
on~\cite[Th.~4.1,~Ch.~1]{rei}) the set of all finite linear
combinations of translates of $h$ are dense in $L^1$, or based
on~\cite[Th.~7.2.9]{rud} that translates of $h$ also span $L^2$,
i.e.\ the space of square integrable functions. Note that versions
for real-valued integrable functions defined on $\R^2$ can be
obtained from the cited theorems as easy exercises, yielding the
corresponding approximation results for real-valued $L^1$-
respectively $L^2$\nobreakdash-functions. Already these simple
observations indicate the potential of the approach for
approximating e.g.\ less regular functions. In the following, we
give a similar result for integrable payoff functions depending
only on two positive prices.
\begin{prop}
  Integrable functions with representation~(\ref{eq:long-version})
  are dense in $L^1(\R^2_+)$.
\end{prop}

\begin{proof}
We extend $f$ to a function $F\in L^1(\R^2)$ by setting
$F(x)=f(x)$ for $x\in\R_+^2$ and $F(x)=0$ otherwise. Further, we
set $h(x,y)=\frac{1}{2\pi}\exp(-\frac{1}{2}(x^2+y^2))$. As a
consequence of Wiener's Approximation Theorem, we have that for
arbitrary $\varepsilon>0$ there exist $n\in \mathbb{N}$,
$\lambda_k$ and $(a_k,b_k)_{k=1}^n$, where $\lambda_k$, $a_k$,
$b_k\in \R$ such that for $H\colon \R^2\to \R$ defined as
\begin{equation*}
H(x,y)=\sum_{k=1}^n \lambda_k h(x-a_k,y-b_k),
\end{equation*}
we obtain
\begin{equation*}
\|H-F\|_{L^1(\R^2)}<\varepsilon.
\end{equation*}
We now define $\tilde{h}_k\colon \R^2_+\to \R$ by
$\tilde{h}_k(x,y)=h(x-a_k,y-b_k)$ and
$\tilde{h}(x,y)=\sum_{k=1}^n\lambda_k\tilde{h}_k(x,y)$. Note that
$\tilde{h}$ is now only defined for $(x,y)\in \R^2_+$, and there
it agrees with $H$. Hence, we can represent the functions $\tilde
h_k$ and also $\tilde{h}$ the way we want. 
Furthermore,
\begin{align*}
\|f-\tilde{h}\|_{L^1(\R^2_+)}=\|F-H\|_{L^1(\R^2_+)}\leq
\|F-H\|_{L^1(\R^2)}<\varepsilon.
\end{align*}
\end{proof}

Other interesting ideas concerning solving practical implementing
problems can be found in other literature about the one
dimensional case, see e.g.~\cite{car:lee08}.

\section*{Acknowledgements}
   The authors are grateful to  Rolf Burgermeister, and Markus
   Liechti for useful hints from practice, to Peter L.\ J{\o}rgensen, Thomas
   Kokholm, and Thorsten Rheinl\"ander
   for theoretical hints, to Gabriel Maresch, Christoph Haberl
   for very helpful discussions, and to Peter Carr for drawing our
   attention to the  problematics in this area. The final special thanks
   go out to Ilya Molchanov for his support in all areas.
   This work was supported by the Swiss National Science Foundation
    Grant Nr.\ 200021-126503 and $\text{PBBEP3}_{-}\text{130157}$.

\newcommand{\noopsort}[1]{} \newcommand{\printfirst}[2]{#1}
  \newcommand{\singleletter}[1]{#1} \newcommand{\switchargs}[2]{#2#1}

\bibliographystyle{alpha}
\bibliography{test}

\end{document}